\documentstyle[12pt]{article}

\def\nn{\nonumber \\}

\def\k{{\not\! k}}
\def\p{{\not\! p}}
\def\del{{\not\! \partial}}
\def\A{{\not\!\! A}}
\def\e{{\rm e}}
\def\a{a}

\advance\oddsidemargin by -\textwidth
\advance\oddsidemargin by -1in
\evensidemargin=\oddsidemargin
\begin{document}

$\mbox{ }$
\vspace{-3cm}
\begin{flushright}
\begin{tabular}{l}
{\bf TIT/HEP-388 }\\
March, 1998
\end{tabular}
\end{flushright}

\vspace{2cm}
\begin{center}
\Large
{\baselineskip26pt \bf 	Non-perturbative solutions\\
                       	   of the SD equation \\
			for an Abelian gauge field theory}
\end{center}
\vspace{1cm}
\begin{center}
\large
{Tsuguo Mogami\footnote{e-mail: mogami@th.phys.titech.ac.jp}}
\end{center}
\normalsize
\begin{center}
{\it Department of Physics, Tokyo Institute of Technology,}\\
{\it Oh-okayama, Meguro, Tokyo 152, Japan}\\
\end{center}
\vspace{2cm}
\begin{center}
\normalsize
ABSTRACT
\end{center}
{\rightskip=2pc 
\leftskip=2pc 
\normalsize
In this article we present a mechanism in which we find new
non-perturbative solutions of quantum electrodynamics in four
dimensions.  Two non-perturbative solutions are found for approximate 
Schwinger-Dyson equations.  
The mass ratio of the three solutions (one is ordinary solution) is 
approximately 
$1:(8\pi/\alpha)^{2/3}:8\pi/\alpha$, which is close to the realistic
mass ratio of $e^-,\mu^-$ and $\tau^-$.
\vglue 0.6cm}

\newpage

In this article, we will find new non-perturbative solutions for
quantum electrodynamics in continuous four dimensional space-time:
\begin{equation}
{\cal L}(x) 
	= -{1\over4}F_{\mu\nu}F^{\nu\mu} + \bar\psi(i\del-e\A-m) \psi,
\label{Lagrangian}
\end{equation}
where $F_{\mu\nu}=\partial_\mu A_\nu-\partial_\nu A_\mu$.  

The Schwinger-Dyson(SD) equation for the electron two point function
$S(p)$ is
\begin{equation}
-i(\p-m) S(p) +ie\int{d^4k\over(2\pi)^4} \ \gamma^\mu S(p-k)
	\Lambda^\nu(p-k,p) S(p)  D_{\mu\nu}(k) = 1,
\label{SD}
\end{equation}
where $D_{\mu\nu}(k)$ is the photon two-point function. $\Lambda^\mu$ is
the 1PI three point function.  The SD equations itself are 
non-perturbative.\cite{SD}  If we could solve the SD 
equations non-perturbatively, we would understand 
non-perturbative phenomena.

We must discuss gauge fixing before solving the equations.
We consider lattice gauge theories to be more natural regularization of 
gauge symmetry, because we need not to fix gauge.  If we force to 
introduce gauge fixing into a lattice gauge theory, only Landau gauge is 
allowed as will be seen below.  Therefore we prefer Landau gauge also in
the continuous theory.

At first, let us recall the derivation of non-Landau gauge, especially 
$R_\xi$-gauge, in continuum.
Following Dirac's prescription for constrained system and Faddeev-
Popov's argument, the path integral formula is
\begin{equation}
\int {\cal D}A {\cal D}\psi {\cal D}\bar\psi
\prod_x \delta(\partial_\mu A^\mu -f(x)) \ \Delta[A]
	\ {\cal O}(x_1,x_2,\cdots) \exp{iS},
\label{8}
\end{equation}
where $S$ is the action and ${\cal O}$ is a operator to be measured.  
The FP(Faddeev-Popov)-determinant $\Delta[A]$ is just a constant since 
we are considering U(1) gauge theory.  If we operate
\begin{equation}
\int {\cal D}f \exp i\lambda^{-1}\int d^4x f^2(x)
\label{Df}
\end{equation}
on eq(\ref{8}), we obtain $R_\xi$-gauge with its constant $\lambda$.
This integration is responsible for fictitious propagation of the 
longitudinal modes.

Let us see if this discussion may be justified in the lattice
regularization.  We consider lattice QED in 4-dimensional Euclidean 
space with its lattice spacing $\a$.  The action is
\begin{equation}
S = \sum_x \a^4 \left\{ 
\sum_\mu (\bar\psi_{x+\hat\mu}U_{x,\mu}
		-\bar\psi_{x-\hat\mu}
	U^\dagger_{x-\hat\mu,\mu})\gamma^\mu \psi_x
+ \bar\psi_x \psi_x 
+ {1\over2e^2}\sum_{\mu\nu} U_{x,\mu} U_{x+\hat\mu,\nu}
		U^\dagger_{x+\hat\nu,\mu} U^\dagger_{x,\nu}
\right\},
\end{equation}
where we denote the unit vector in $\mu$ direction by $\hat\mu$, and finite
constants on each term are omitted.  Roughly speaking, the link variable
$U_{x,\mu}$ is related to the gauge field $A^\mu(x)$ by $U_{x,\mu}=\a^{-1}
\exp{iea A^\mu(x)}$.  We quantize lattice theory by just doing
path-integration without fixing gauge:
\begin{equation}
\int \prod_{x,\mu} dU_{x,\mu} \prod_x d\psi_x d\bar\psi_x 
	\ {\cal O}(x_1,x_2,\cdots) \ {\rm e}^{-S}.
\label{lattice}
\end{equation}
(Readers should not be confused by using the same notation $S,\psi,x$,
etc. as the continuous theory.)  This expression is equivalent to
\begin{equation}
\left.\int \prod_{x,\mu} dU_{x,\mu} \prod_x d\psi_x d\bar\psi_x 
	\Delta[U] \prod_x\delta({\rm Re}\Delta_\mu U_{x,\mu} - f_x) 
	\ {\cal O}(x_1,x_2,\cdots)\ {\rm e}^{-S}  \right. ,
\label{gauge_fixed_lattice}
\end{equation}
where $\Delta_\mu f_x \equiv (f_{x+\mu}-f_x)/\a$ and $\Delta[U]$ is
defined as
\begin{equation}
(\Delta[U])^{-1} \equiv \int \prod_x d\theta_x 
	\ \delta({\rm Re}\Delta_\mu 
(\e^{i\theta_{x+\mu}} U_{x,\mu}\e^{-i\theta_x}) - f_x).
\label{lattice_fp_det}
\end{equation}  

If we could operate $\prod_x \int df_x \exp \alpha\a^4\sum_x f_x^2$ on 
eq.(\ref{gauge_fixed_lattice}), we would obtain $R_\xi$-gauge.
Eq.(\ref{gauge_fixed_lattice}) is equivalent to eq.(\ref{lattice}) as 
far as
${\rm Re}\Delta_\mu U_{x,\mu}^\theta - f_x = 0$ has a solution.
$U_{x,\mu}$ is as large as $1/\a$ at most and $\Delta_\mu U_{x,\mu}$
is $1/\a^2$ at most.  Thus the equation has solution only when $|f_x|$ 
is smaller than $1/\a^2$.  However, the $\int df_x \exp{i\alpha\a^4
f_x^2}$ integration goes out of this range even if $\a \to 0$.  
Therefore generic $R_\xi$-gauge is not justified in lattice gauge theory.  

This argument may also be heuristically applied to continuous QED.  
Suppose the
continuous theory has momentum cutoff $1/a$, which implies minimum
distance $a$.  The essence of electromagnetic force is the change of
the phase of electrons by the gauge field $A^\mu$.  The change is
$A^\mu \times {\rm (length)}$.  Any change of phase larger than $2\pi$
at minimum distance does not make sense.  This means that $|A^\mu|$ must
be smaller than $2\pi/a$ and $\partial_\mu A^\mu$ must be smaller than
$2\pi/a^2$.  Therefore the ${\cal D}f$ integration in eq.(\ref{Df}) is not
justified.  We see that $R_\xi$-gauge comes from ill-definedness of 
cutoff in continuum theory.

\section*{The first non-perturbative solution}

Now let us approximate and solve the SD equations.  At first, expand the
three point function in eq.(\ref{SD}), and we get:
\begin{eqnarray}
&&\biggl\{-i(\p-m)
+e^2\int{d^4k\over(2\pi)^4}
	\ \gamma^\mu S(p-k)\gamma^\nu \ D_{\mu\nu}(k)		\nn
&&-e^4\int{d^4k_1 d^4k_2\over(2\pi)^8}
\ \gamma^\rho S(p-k_2)\gamma^\mu S(p-k_1-k_2)\gamma^\sigma S(p-k_1)
\gamma^\nu 
	\ D_{\rho\sigma}(k_2) D_{\mu\nu}(k_1) 			\nn
&&+\cdots\biggr\}S(p) = 1.
\label{expansion}
\end{eqnarray}

Here we take up to the one-loop term.  We also assume that the electric 
charge remains small even with non-perturbative effect.  Then $D^{\mu
\nu}(k)$ gets only small radiative correction.  Therefore we may 
approximate $D^{\mu\nu}$ by the bare photon propagator $D_0^{\mu\nu}(k) 
\equiv {-i\over k^2 +i\epsilon} (g^{\mu\nu} - {k\mu k\nu\over k^2})$.  
Only the electron two point function is not substituted by the bare one.
The mass term will be incorporated later.
With these approximation, equation(\ref{SD}) is simply
\begin{equation}
\left\{-i\p
+e^2 \int{d^4k\over(2\pi)^4}\gamma_\mu S(p-k)\gamma_\nu
 \ D_0^{\mu\nu}(k)\right\}S(p) = 1.
\label{rSD}
\end{equation}
Let us denote its solution by $S_0(p)$, since it is an approximate 
solution.  The solution of this equation is $S_0(p)=1/c\p$, where $c$ is
a constant to be determined.

The second term in eq.(\ref{rSD}) is calculated as follows.  Using
\begin{equation}
\Delta_F(x) 
\equiv \int {d^4k \over (2\pi)^4} {i \e^{-ikx} \over k^2+i\epsilon}
= -{1 \over 4\pi^2 (x^2-i\epsilon)},
\end{equation}
1-loop 1PI two-point function in the real space is
\begin{equation}
\Sigma(x) = -e^2 \gamma_\mu S(x) \gamma_\nu \ D_0^{\mu\nu}(x)
= -e^2 \gamma_\mu (i\del \Delta_F(x)) \gamma_\nu 
\ \{-g^{\mu\nu}+{1\over 2}\partial^\mu(x^\nu\Delta_F(x)) \}.
\end{equation}
This $\Sigma(x)$ should be treated as a hyperfuntion.  With an arbitrary
function $f(x)$
\begin{equation}
\int d^4x \ f(x)\Sigma(x) 
= \int d^4x f(x) {e^2 \over 2} (i\del^2\Delta_F(x))	
     \gamma_\nu x^\nu \Delta_F(x)
= \int d^4x f(x) {\alpha \over 8\pi^2} \del\delta^4(x),
\end{equation}
which means that $\Sigma(x)=\alpha\del\delta^4(x)/{8\pi}$.

Now eq.(\ref{rSD}) reduces to an equation for $c$:
\begin{equation}
\left\{ -i\p + {\alpha\over8\pi c}i\p\right\} {i\over c\p+i\epsilon} = 1.
\label{rSDa}
\end{equation}
The solution of this equation is $c=(1\pm\sqrt{1-\alpha/{2\pi}})/2$.
If we take the plus sign, we get a solution which substitutes a
perturbative series of Feynman graphs.  Taking the minus sign, we obtain
an unfamiliar non-perturbative solution: $c \sim \alpha/{8\pi}$. 

What does this new solution mean physically?  The kinetic term of the 
effective action is $\bar\psi S_0^{-1}(p) \psi$.  If we add the mass 
term to the bare Lagrangian, the effective action becomes $-i\bar\psi(
c\p-m)\psi$.  This means that the mass of the new particle\footnote{
This solution does not allow both the new particle and the ordinary
particle at once as far.  We assume multiple fermion fields with the 
same bare mass and charge.  This theory only determines a possible 
physical mass ratio of these fermions.} is $M=m/c$.  
It is interesting to note that, letting $m$ to be electron mass$\sim 
.5110$MeV, $m/c$ is about 1760MeV, which is close to the $\tau$-particle
mass$\sim 1777$MeV.

The Ward-Takahashi(WT) identities guarantee that its charge is same as 
electron.  The WT identities will never be broken even in a
non-perturbative solution as far as the U(1) gauge symmetry is not 
broken.   If it were bronken, we would have a effectively strongly 
coupled theory and 
would have difficulty in radiative correction.\footnote{Rochev and 
Saponov presented similar mechanism in \cite{RS} for the scalar 
$\phi^4$ theory.  Because the theory do not have WT-identities, the
new particles are strongly coupled.  Such theory does not have virtue of
pseudo perturbation seen in our theory.}

One may wonder why we did not substitute $-iec\gamma^\mu$ for 
$\Lambda^\mu$ in eq.(\ref{SD}) presuming the WT identity.  If we did so, 
the contribution summed in this
approximation is asymmetric in the change of direction of electron line.
This approximation bears only perturbative solution.  The asymmetry
might have destroyed the non-perturbative solution.

To summarize, we have effective Feynman rules in which the fermion
propagator is $S_0(p)=i/(c\p-m)$ and the three point vertex
is $-iec\gamma^\mu$.  Though $S_0$ consists of infinite sum of diagrams,
we may regard it as the bare propagator of the new particle.  

\section*{Radiative Corrections and Renormalization}

The tree level amplitudes can be read from above effective Feynman 
rules.  Now we discuss how radiative corrections and renormalization 
should be made for one-loop.  The discussion will also clarify how the 
contributions ignored above are incorporated.  

Using the effective Feynman rules, the effective one-loop correction is
\begin{equation}
 -c^2 e^2\int{d^4k \over (2\pi)^4} 
	\ \gamma_\mu {i\over c(\p-\k)-m}
	\gamma_\nu D_0^{\mu\nu}(k).
\label{1loop}
\end{equation}
The $\p$ proportional part in this one-loop 1PI two-point function is 
independent of the mass.  The diagrams calculated by effective Feynman 
rules include sum of higher order contributions of true bare 
propagators.  If we expand eq.(\ref{1loop}) in terms of the bare 
propagator, the first $\p$ proportional term in this sum is already 
taken in 
eq.(\ref{rSD}).  Then we must not take $\p$ part for the first order 
radiative correction.  

Contrary, the $m$ proportional part of expression (\ref{1loop}) is not 
taken into eq.(\ref{rSD}) at all.
The effective action gets this correction.  Its mass term gets infinite 
contributon $(3\alpha/4\pi)m\log\Lambda^2$, where $\Lambda$ is the 
cutoff.  Does this mean that the physical mass of the heavier particle is 
affected by the cutoff?  Now we must recall renormalization.  The bare 
mass should get the counter term: $-(3\alpha/4\pi)m\log\Lambda^2$, 
because we defined the physical mass of the perturbative solution to be 
$m$.  Therefore $\Lambda$ is cancelled.
This is an example of compatibility of this theory with renormalization.
Moreover, this radiative correction agrees with unitarity.
The radiative corrections from the higher order diagrams may be done in 
the same way.  All the contributions omitted from eq.(\ref{rSD}) may be 
incorporated order by order as radiative corrections. 

Methods to obtain ``non-perturbative'' solution often  have no room to 
incorporate radiative corrections.  In seeking non-perturbative 
solutions, neglecting tiny contributions in the equation may cause us 
to invent a ghost solution because of non-perturbative nature.  This 
theory takes care of all the contributions at least pseudo-
perturbatively.

\section*{Further discussion}

What shall we find if we take up to the two-loop in eq.(\ref{expansion})?
To derive the two loop contribution, we substitute
\begin{equation}
\Lambda^\mu(p',p) = -ie\gamma^\mu 
-ie^3\int{d^4k \over(2\pi)^4}
\ \gamma^\rho S(p'-k)\gamma^\mu S(p-k)\gamma^\sigma
	D_{\rho\sigma}(k) 
\label{expandL}
\end{equation}
into eq.(\ref{SD}).  The one-loop term in this expression has non-local 
contributions and it makes the equation unsolvable.  Let us recall that, 
in the case of two point 1PI function we only had local part.
Here we assume to take local part of eq(\ref{expandL}) 
corresponding to the two point function, i.e. 
$\Lambda^\mu = -ie\gamma^\mu+ie \alpha \gamma^\mu /{8\pi}c^2$, where
the second term is calculated by applying formal WT identiry to 
$S(p)=i/c\p$.  With these approximation, the SD equation is
\begin{equation}
\left\{
-i\p+{\alpha\over8\pi c}i\p -{\alpha^2\over(8\pi)^2 {c}^3}i\p 
(1-{\alpha\over 8\pi c})^4
\right\} {i\over c\p}= 1. 
\end{equation}
The extra factor $(1-{\alpha / 8\pi c})^4$ comes from the one-loop 
level correction to each vertex.
Besides the first solution, this equation has two new solution 
$c\sim 1/214$ and $c\sim 1/3660$.  Mass of the former particle 
is about 109MeV.  This is not so far from $\mu$-particle mass$\sim 
105.7$MeV.  The other solution may be a ghost solution.


Readers may wonder what happens in the cases of more loops caluculation.
We don't have definite answer on this question.

We must admit that mathematical validity of this theory is not so 
evident.  Indeed, such exotic solutions are false in many cases, but 
sometimes have profound meaning like negative energy solution found by 
Dirac lead to anti-matter. 
Rather good agreement of mass hierarchy with real leptons suggests that 
this method at least may reflect the essence of true theory to be sought.  

\section*{Acknowledgements}
The author is supported by JSPS Reserch Fellowship for Young Scientists.
The author is grateful to professor N. Kawamoto and Y. Watabiki 
for discussions.


\vfill

\end{document}